\documentclass[aps,prd,nofootinbib,floatfix,superscriptaddress,secnumarabic]{revtex4}
\usepackage{amsmath}
\usepackage{graphicx}
\usepackage{xcolor}
\newcommand{\MSbar}{{\overline{\rm MS}}}
\newcommand{\pa}{\partial}
\newcommand{\gtilde}{\frac{g^2}{16 \, \pi^2}\; }

\newcommand{\la}{\lambda}

\newcommand{\be}{\begin{equation}}
\newcommand{\ee}{\end{equation}}
\newcommand{\bea}{\begin{eqnarray}}
\newcommand{\eea}{\end{eqnarray}}
\def\lsim{\mathrel{\rlap{\lower4pt\hbox{\hskip1pt$\sim$}}
    \raise1pt\hbox{$<$}}}                % less than or approx. symbol

\begin{document}

\title{Supersymmetric QCD on the Lattice: Fine-Tuning of the Yukawa Couplings}

\author{M.~Costa}
\email{kosta.marios@ucy.ac.cy}
\affiliation{Department of Physics, University of Cyprus, Nicosia, CY-1678, Cyprus}
\affiliation{Department of Chemical Engineering, Cyprus University of Technology, Limassol, CY-3036, Cyprus}
%Department of Chemical Engineering, Cyprus University of Technology, 30 Archbishop Kyprianou Str., 3036, Limassol, Cyprus

\author{H.~Herodotou}
\email{herodotos.herodotou@ucy.ac.cy}
\affiliation{Department of Physics, University of Cyprus, Nicosia, CY-1678, Cyprus}

\begin{abstract}
We determine the fine-tuning of the Yukawa couplings of supersymmetric QCD, discretized on a lattice. We use perturbation theory at one-loop level. The Modified Minimal Subtraction scheme ($\overline{{\rm MS}}$) is employed; by its definition, this scheme requires perturbative calculations, in the continuum and/or on the lattice. On the lattice, we utilize the Wilson formulation for gluon, quark and  gluino fields; for squark fields we use na\"ive discretization. The sheer difficulties of this study lie in the fact that different components of squark fields mix among themselves at the quantum level and the action's symmetries, such as parity and charge conjugation, allow an additional Yukawa coupling. Consequently, for an appropriate fine-tuning of the Yukawa terms, these mixings must be taken into account in the renormalization conditions. All Green's functions and renormalization factors are analytic expressions depending on the number of colors, $N_c$, the number of flavors, $N_f$, and the gauge parameter, $\alpha$, which are left unspecified. Knowledge of these renormalization factors is necessary in order to relate numerical results, coming from nonperturbative studies, to the renormalized, ``physical" Green's functions of the theory.
\end{abstract}

\maketitle

\section{Introduction}

Over the past decades, supersymmetry (SUSY) has been considered a prime candidate for resolving a number of open problems related to the Standard Model (SM), such as the candidates to explain the nature of dark matter~\cite{Bertone:2004pz}, the unification of the electromagnetic, weak and strong forces suggested by Grand Unified Theory (GUT)~\cite{Susskind:1982mw, Ellis:1983ew}, and the hierarchy problem~\cite{Susskind:1982mw}. Unbroken SUSY dictates equal fermionic and bosonic degrees of freedom within supermultiplets. However, SUSY particles have remained elusive~\cite{ParticleDataGroup:2022pth}, necessitating the nonperturbative study of the SUSY phase transition. Supersymmetric models of strongly coupled theories are a very promising models for new physics Beyond the SM and lattice investigations of supersymmetric extensions of QCD are becoming  within  reach. However, there are several well-known obstacles arising from the breaking of SUSY in a regularized theory on the lattice~\cite{Schaich:2022xgy}, including the necessity for fine tuning of the theory's bare Lagrangian~\cite{Giedt:2009yd, Schaich:2014, Bergner:2016sbv}. 

%Another important motivation for the nonperturbative investigations of SUSY theories are theoretical conjectures about the confinement mechanism and relations to Gauge/Gravity duality. 
An additional significant incentive for delving into nonperturbative explorations of supersymmetric theories stems from theoretical conjectures concerning confinement mechanisms and their connections to Gauge/Gravity duality. These have their foundations in the enhanced symmetries of supersymmetric gauge theories and it would be interesting to extend and relate them to QCD or Yang-Mills theory. This requires more general insights into the nonperturbative regime of supersymmetric theories. Numerical lattice simulations would be an ideal nonperturbative first-principles tool to investigate gauge theories with SUSY. However, it is unavoidable to break SUSY in any non-trivial theory on the lattice. In general, fine tuning is required to restore supersymmetry in the continuum limit (see, e.g., Ref.~\cite{Athron:2007ry}), which can be guided by signals provided by the SUSY Ward identities~\cite{Montvay:2002, Ali:2018fbq}. The analysis of SUSY Ward identities requires the renormalization of the supercurrent~\cite{Taniguchi:1999fc}, which can mix due to broken supersymmetry with other operators of the same or lower dimension. Even though lattice breaks ${\cal{N}}=1$ supersymmetry explicitly~\cite{Curci:1986sm}, it is the best method at present to obtain  quantitative results. There are also other theories with extended supersymmetry~\cite{Sugino:2008yp, Catterall:2009it, Giedt:2014}, which preserve some supercharges on the lattice; however in this work we focus in ${\cal{N}}=1$ supersymmetric QCD (SQCD) which is more realistic in the sense that it is directly related to extensions of the SM. 

The gauge invariance of the lattice SQCD action dictates that some of the action's interaction terms will share the same coupling constant, $g$ (gauge coupling). This is particularly applicable to the kinematic terms containing covariant derivatives, resulting in gluons coupling with quarks, squarks, gluinos, and other gluons, all governed by the same gauge coupling constant. The Yukawa interactions involving quarks, squarks, and gluinos, as well as the four-squark interactions, have the potential to feature distinct couplings, at the quantum level. Furthermore, new terms may also emerge, necessitating careful fine-tuning on the lattice. By exploiting the symmetries of the Wilson lattice action, we can predict these potentially novel interaction terms. However, with the actual computation we can understand if they will arise at the quantum level, and more importantly we can determine their renormalizations to certain perturbative order. 

In this article, we present one-loop perturbative results regarding the renormalization of the Yukawa couplings, which are obtained by using the ${\cal{N}} =1$ Supersymmetric QCD with the gauge group $SU(N_c)$ and $N_f$ flavors. After presenting the basics of the computation setup (Section \ref{comsetUP}), we start with a discussion of the renormalization of the Yukawa couplings (Section \ref{couplingY}) both in dimensional and lattice regularizations. We utilize the $\MSbar$ renormalization scheme and we determine the renormalization factors to one-loop order. Finally, outlook and future plans are briefly outlined (Section~\ref{summary}). We also provide an appendix (Appendix~\ref{majoranaS}) that elaborates the Majorana nature of the gluino field within the functional integral framework.

\section{Formulation and Computational Setup}
\label{comsetUP} 

In this section we shortly introduce the computational setup of our study, along with the notation used in the paper. We give the definitions for the symmetries of the action as well as the transformation properties of the Yukawa terms. These symmetries allow an additional linear combination of ``Yukawa type'' operators, which can in principle appear at the quantum level. In addition, we provide the Feynman diagrams for the calculation of the three-point (3-pt) Green's functions which we must compute in order to extract the renormalization of the Yukawa couplings. Several prescriptions for defining $\gamma_5$ in $D$ dimensions~\cite{Jones:1983ip} are also presented in the end of this section. Since $\MSbar$ renormalized Green's functions are computed in Dimensional Regularization ($DR$) we have also introduced the continuum action of SQCD. In $DR$ the regulator, $\epsilon$, is defined by $D \equiv 4 - 2 \epsilon$; in the Lattice Regularization ($LR$) the lattice spacing, $a$, serves as regulator for the $UV$ divergences. 

In the Wess-Zumino (WZ) gauge, the SQCD action contains the following fields: the gluon together with the gluino; and for each quark flavor, a Dirac fermion (quark) and two components of squarks. Although the action of SQCD used in this calculation can be found in the literature, e.g. in Ref.~\cite{Wess:1992cp, Martin:1997ns}, for completeness' sake we present it here; in the continuum and in Minkowski space, the action of SQCD is:
\bea
{\cal S}_{\rm SQCD} & = & \int d^4x \Big[ -\frac{1}{4}u_{\mu \nu}^{\alpha} {u^{\mu \nu}}^{\alpha} + \frac{i}{2} \bar \lambda^{\alpha}  \gamma^\mu {\cal{D}}_\mu\lambda^{\alpha}  \nonumber \\
&-& {\cal{D}}_\mu A_+^{\dagger}{\cal{D}}^\mu A_+ - {\cal{D}}_\mu A_- {\cal{D}}^\mu A_-^{\dagger}+ i \bar \psi  \gamma^\mu {\cal{D}}_\mu \psi  \nonumber \\
&-&i \sqrt2 g \big( A^{\dagger}_+ \bar{\lambda}^{\alpha}  T^{\alpha} P_+ \psi   -  \bar{\psi}  P_- \lambda^{\alpha}   T^{\alpha} A_+ +  A_- \bar{\lambda}^{\alpha}  T^{\alpha} P_- \psi   -  \bar{\psi}  P_+ \lambda^{\alpha}   T^{\alpha} A_-^{\dagger}\big)\nonumber\\  
&-& \frac{1}{2} g^2 (A^{\dagger}_+ T^{\alpha} A_+ -  A_- T^{\alpha} A^{\dagger}_-)^2 + m ( \bar \psi  \psi  - m A^{\dagger}_+ A_+  - m A_- A^{\dagger}_-)\Big] \,,
\label{susylagr}
\eea
where $\psi$ ($A_\pm$) is the quark field (squark field components), $u_{\mu} = u_{\mu}^\alpha \, T^\alpha$ ($\lambda = \lambda^\alpha \,
T^\alpha$) is the gluon (gluino) field; $T^{\alpha}$ are the generators of the $SU(N_c)$ gauge group and $P_\pm= (1 \pm \,\gamma_5)/2$ are projectors. Quarks and squarks should also be assigned with color indices in the fundamental representation of the gauge group $SU(N_c)$, whereas gluons and gluinos carry an $\alpha$ index which is a color index in the adjoint representation of the gauge group. The definitions of the covariant derivatives and of the gluon field tensor are:
\bea
{\cal{D}}_\mu A_+ &=&  \pa_{\mu} A_+ + i g\,u_{\mu}^{\alpha}\,T^{\alpha}\,A_+ \nonumber \\
{\cal{D}}_\mu A_-^{\dagger} &=&  \pa_{\mu} A_-^{\dagger} + i g\,u_{\mu}^{\alpha}\,T^{\alpha}\,A_-^{\dagger} \nonumber \\
{\cal{D}}_\mu A_- &=&  \pa_{\mu} A_- - i g\,A_-\,T^{\alpha}\,u_{\mu}^{\alpha} \nonumber \\
{\cal{D}}_\mu A_+^{\dagger} &=&  \pa_{\mu} A_+^{\dagger} - i g\,A_+^{\dagger}T^{\alpha}\,u_{\mu}^{\alpha} \nonumber \\
{\cal{D}}_\mu \psi &=&  \pa_{\mu} \psi+ i g\,u_{\mu}^{\alpha} \,T^{\alpha}\,\psi \nonumber\\
%{\cal{D}}_\mu \psi_+^{\dagger} &=&  \pa_{\mu} \psi_+^{\dagger} - i g \,\psi_+^{\dagger} \,T^{\alpha}\,u_{\mu}^{\alpha} \nonumber \\
{\cal{D}}_\mu \lambda &=&  \pa_{\mu} \lambda + i g \,[u_{\mu},\lambda] \nonumber\\
 u_{\mu \nu} &=& \pa_{\mu}u_{\nu} - \pa_{\nu}u_{\mu} + i g\, [u_{\mu},u_{\nu}]. 
\eea
The parts of the continuum and lattice SQCD actions that are associated with the quark and the squark fields (Eqs.~(\ref{susylagr}) and~(\ref{susylagrLattice}), respectively) involve a summation over flavor indices\footnote{A double summation over flavors is implicit in 
the 4-squark term of the action (last line of Eqs.~(\ref{susylagr}) and~(\ref{susylagrLattice})).};  these flavor indices are implicit within our expressions.

The above action in Eq.~(\ref{susylagr}) is invariant under these supersymmetric transformations with a Majorana Grassmann parameter $\xi$:
\bea
\delta_\xi A_+ & = & - \sqrt2  \bar\xi  P_+\psi  \, , \nonumber \\
\delta_\xi A_- & = & - \sqrt2  \bar\psi  P_+ \xi   \, , \nonumber \\
\delta_\xi (P_+ \psi_{D}) & = & i \sqrt2 ({\cal{D}}_\mu A_+) P_+ \gamma^\mu \xi   - \sqrt2 m P_+ \xi  A_-^{\dagger}\, , \nonumber \\
\delta_\xi (P_- \psi ) & = &  i \sqrt2 ({\cal{D}}_\mu A_-)^{\dagger} P_- \gamma^\mu \xi   - \sqrt2 m  A_+ P_- \xi \, ,\nonumber \\
\delta_\xi u_{\mu}^{\alpha} & = & -i \bar \xi  \gamma_\mu \lambda^{\alpha} , \nonumber \\
\delta_\xi \lambda^{\alpha}  & = & \frac{1}{4} u_{\mu \nu}^{\alpha} [\gamma^{\mu},\gamma^{\nu}] \xi  - i g \gamma^5 \xi  (A^{\dagger}_+ T^{\alpha} A_+ -  A_- T^{\alpha} A^{\dagger}_-)\,.
\label{susytransfDirac}
\eea

As in the case with the quantization of ordinary gauge theories, additional infinities will appear upon functionally integrating over gauge orbits. The standard remedy is to introduce a gauge-fixing term in the Lagrangian, along with a compensating Faddeev-Popov ghost term. The resulting Lagrangian, though no longer gauge invariant, is still invariant under BRST transformations. This procedure of gauge fixing guarantees that Green's functions of gauge invariant objects will be gauge independent to all orders in perturbation theory. We use the ordinary gauge fixing term and ghost contribution arising from the Faddeev-Popov gauge fixing procedure:
\begin{equation}
{\cal S}_{GF}= \frac{1}{\alpha}\int d^4x {\rm{Tr}} \left( \partial^\mu u_\mu\right)^2,
\label{sgf}
\end{equation}
where $\alpha$ is the gauge parameter ($\alpha=1(0)$ corresponds to Feynman (Landau) gauge), and 
\begin{equation}
{\cal S}_{Ghost}= - 2 \int d^4x {\rm{Tr}} \left( \bar{c}\, \partial^{\mu}D_\mu  c\right),  \quad {\cal{D}}_\mu c =  \pa_{\mu} c -i g \,[u_\mu,c],
\label{sghost}
\end{equation}
where the ghost field, $c$, is a Grassmann scalar which transforms in the adjoint representation of the gauge group. This gauge fixing term breaks supersymmetry. However, given that the renormalized theory does not depend on the choice of a gauge fixing term, and given that both dimensional and lattice regularizations violate SUSY at intermediate steps, one may choose this standard covariant gauge fixing term.

In Refs.~\cite{Costa:2017rht} and~\cite{Costa:2018mvb}, the first lattice perturbative computations in the context of SQCD were presented; apart from the Yukawa and the quartic couplings~\cite{Herodotou:2022xhz}, ~\cite{Herodotou:2023xhz}, we have extracted the renormalization of all parameters and fields appearing in Eq.~(\ref{susylagrLattice}) using Wilson gluons and fermions.  The results in  references~\cite{Costa:2017rht,Costa:2018mvb} will find further use in the present work. 

In our lattice calculation, we extend Wilson's formulation of the QCD action, to encompass SUSY partner fields as well. In this standard discretization quarks, squarks and gluinos live on the lattice sites, and gluons live on the links of the lattice: $U_\mu (x) = e^{i g a T^{\alpha} u_\mu^\alpha (x+a\hat{\mu}/2)}$; $\alpha$ is a color index in the adjoint representation of the gauge group. This formulation leaves no SUSY generators intact, and it also breaks chiral symmetry; hence, the need for fine-tuning will arise in numerical simulations of SQCD. For Wilson-type quarks and gluinos, the Euclidean action ${\cal S}^{L}_{\rm SQCD}$ on the lattice becomes:       
\bea
{\cal S}^{L}_{\rm SQCD} & = & a^4 \sum_x \Big[ \frac{N_c}{g^2} \sum_{\mu,\,\nu}\left(1-\frac{1}{N_c}\, {\rm Tr} U_{\mu\,\nu} \right ) + \sum_{\mu} {\rm Tr} \left(\bar \lambda  \gamma_\mu {\cal{D}}_\mu\lambda  \right ) - a \frac{r}{2} {\rm Tr}\left(\bar \lambda   {\cal{D}}^2 \lambda  \right) \nonumber \\ 
&+&\sum_{\mu}\left( {\cal{D}}_\mu A_+^{\dagger}{\cal{D}}_\mu A_+ + {\cal{D}}_\mu A_- {\cal{D}}_\mu A_-^{\dagger}+ \bar \psi  \gamma_\mu {\cal{D}}_\mu \psi  \right) - a \frac{r}{2} \bar \psi   {\cal{D}}^2 \psi  \nonumber \\
&+&i \sqrt2 g \big( A^{\dagger}_+ \bar{\lambda}^{\alpha}  T^{\alpha} P_+ \psi   -  \bar{\psi}  P_- \lambda^{\alpha}   T^{\alpha} A_+ +  A_- \bar{\lambda}^{\alpha}  T^{\alpha} P_- \psi   -  \bar{\psi}  P_+ \lambda^{\alpha}   T^{\alpha} A_-^{\dagger}\big)\nonumber\\  
&+& \frac{1}{2} g^2 (A^{\dagger}_+ T^{\alpha} A_+ -  A_- T^{\alpha} A^{\dagger}_-)^2 - m ( \bar \psi  \psi  - m A^{\dagger}_+ A_+  - m A_- A^{\dagger}_-)
\Big] \,,
\label{susylagrLattice}
\eea
where:  $U_{\mu \nu}(x) =U_\mu(x)U_\nu(x+a\hat\mu)U^\dagger_\mu(x+a\hat\nu)U_\nu^\dagger(x)$, and a summation over flavors is understood in the last three lines of Eq.~(\ref{susylagrLattice}). The 4-vector $x$ is restricted to the values $x = n a$, with $n$ being an integer 4-vector. The terms proportional to the Wilson parameter, $r$, eliminate the problem of fermion doubling, at the expense of breaking chiral invariance. In the limit $a \to 0$ the lattice action reproduces the continuum (Euclidean) one. As we will describe below, the bare coupling for the Yukawa terms (third line of Eq.\eqref{susylagrLattice}) need not coincide with the gauge coupling $g$; this requirement will be imposed on the respective renormalized value.

The definitions of the covariant derivatives are as follows:
\bea
{\cal{D}}_\mu\lambda (x) &\equiv& \frac{1}{2a} \Big[ U_\mu (x) \lambda  (x + a \hat{\mu}) U_\mu^\dagger (x) - U_\mu^\dagger (x - a \hat{\mu}) \lambda  (x - a \hat{\mu}) U_\mu(x - a \hat{\mu}) \Big] \\
{\cal D}^2 \lambda (x) &\equiv& \frac{1}{a^2} \sum_\mu \Big[ U_\mu (x)  \lambda  (x + a \hat{\mu}) U_\mu^\dagger (x)  - 2 \lambda (x) +  U_\mu^\dagger (x - a \hat{\mu}) \lambda  (x - a \hat{\mu}) U_\mu(x - a \hat{\mu})\Big]\\
{\cal{D}}_\mu \psi (x) &\equiv& \frac{1}{2a}\Big[ U_\mu (x) \psi  (x + a \hat{\mu})  - U_\mu^\dagger (x - a \hat{\mu}) \psi  (x - a \hat{\mu})\Big]\\  
{\cal D}^2 \psi (x) &\equiv& \frac{1}{a^2} \sum_\mu \Big[U_\mu (x) \psi  (x + a \hat{\mu})  - 2 \psi (x) +  U_\mu^\dagger (x - a \hat{\mu}) \psi  (x - a \hat{\mu})\Big]\\
\label{DAplus}
{\cal{D}}_\mu A_+(x) &\equiv& \frac{1}{a} \Big[  U_\mu (x) A_+(x + a \hat{\mu}) - A_+(x)   \Big]\\
{\cal{D}}_\mu A_+^{\dagger}(x) &\equiv& \frac{1}{a} \Big[A_+^{\dagger}(x + a \hat{\mu}) U_\mu^{\dagger}(x)  -  A_+^\dagger(x)\Big]\\
{\cal{D}}_\mu A_-(x) &\equiv& \frac{1}{a} \Big[A_-(x + a \hat{\mu}) U_\mu^{\dagger}(x)  -  A_-(x)\Big]\\
\label{DAminusdagger}
{\cal{D}}_\mu A_-^{\dagger}(x) &\equiv& \frac{1}{a} \Big[U_\mu (x) A_-^{\dagger}(x + a \hat{\mu})   -  A_-^{\dagger}(x) \Big]
\eea
In Eqs.~(\ref{DAplus})-(\ref{DAminusdagger}) in order to avoid a ``doubling'' problem for squarks we do not use the symmetric derivative; note, however,
that the symmetries of the action are the same for both types of derivatives. A discrete version of a gauge-fixing term, together with the compensating ghost field term, must be added to the action, in order to avoid divergences from the integration over gauge orbits; these terms are the same as in the non-supersymmetric case. In addition, a standard ``measure'' term must be added to the action, in order to account for the Jacobian in the change of integration variables: $U_\mu \to u_\mu$\,. 

In our previous works \cite{Costa:2020keq, Costa:2021pfu, Bergner:2022wnb}, we studied the mixing of certain composite operators upon renormalization. The symmetries of the action play a crucial role to identify the candidate mixing operators. Similarly, in this work, we examine the transformation properties of Yukawa-type operators (gauge-invariant operators of dimension-four, composed of one gluino, one quark, and one squark field) under both parity $\cal{P}$ and charge conjugation $\cal{C}$, and we have determined which specific linear combinations of them remain unchanged. All potential Yukawa terms and their transformation properties are detailed in Table \ref{tb:Ycoupling}.

The symmetries of the lattice action and their definitions are presented below.
\be
{\cal{P}}:\left \{\begin{array}{ll}
&\hspace{-.3cm} U_0(x)\rightarrow U_0(x_{\cal{P}})\, ,\qquad U_k(x)\rightarrow U_k^{\dagger}(x_{\cal{P}}-a\hat{k})\, ,\qquad k=1,2,3\\[4pt]
&\hspace{-.3cm} \psi(x)\rightarrow \gamma_0  \psi(x_{\cal{P}})\\[4pt]
&\hspace{-.3cm}\bar{ \psi}(x) \rightarrow\bar{ \psi}(x_{\cal{P}})\gamma_0\\[4pt]
&\hspace{-.3cm} \la^{\alpha}(x) \rightarrow \gamma_0  \la^{\alpha}(x_{\cal{P}})\\[4pt]
&\hspace{-.3cm}\bar{ \la}^{\alpha}(x) \rightarrow\bar{ \la}^{\alpha}(x_{\cal{P}})\gamma_0\\[4pt]
&\hspace{-.3cm} A_\pm(x) \rightarrow A_\mp^\dagger(x_{\cal{P}})\\[4pt]
&\hspace{-.3cm} A_\pm^\dagger(x) \rightarrow A_\mp(x_{\cal{P}})
\end{array}\right .
\label{Parity}
\ee
where $x_{\cal{P}}=(-{\bf{x}},x_0)$.

\be
{\mathcal {C}}:\left \{\begin{array}{ll}
&\hspace{-.3cm}U_\mu(x)\rightarrow U_\mu^{\star}(x)\, ,\quad \mu=0,1,2,3\\[4pt]
&\hspace{-.3cm}\psi(x)\rightarrow -C \bar{ \psi}(x)^{T}\\[4pt]
&\hspace{-.3cm}\bar{\psi}(x)\rightarrow{\psi}(x)^{T}C^{\dagger}\\[4pt]
&\hspace{-.3cm} \la(x) \rightarrow C \bar{\la}(x)^{T}\\[4pt]
&\hspace{-.3cm}\bar{\la}(x) \rightarrow -{\la}(x)^{T}C^{\dagger}\\[4pt]
&\hspace{-.3cm}A_\pm(x) \rightarrow A_\mp(x) \\[4pt]
&\hspace{-.3cm}A_\pm^\dagger(x) \rightarrow A_\mp^\dagger(x)
\end{array}\right . 
\label{Chargeconjugation}
\ee
where $^{\,T}$ means transpose (also in the $SU(N_c)$ generators implicit in the gluino fields). The matrix $C$ satisfies: $(C \gamma_{\mu})^{T}= C \gamma_{\mu}$, $C^T=-C$ and $C^{\dagger} C=1$. In four dimensions, in a standard basis for $\gamma$ matrices, in which $\gamma_0,\ \gamma_2$ ($\gamma_1,\ \gamma_3$) are symmetric (antisymmetric), $C = - {\rm i} \gamma_0 \gamma_2$. Note that all operators considered in Table \ref{tb:Ycoupling} are flavor singlets. 

\begin{table}[ht]
\begin{center}

  \begin{tabular}{ c | c | c}
\hline \hline
Operators & $\cal{C}$ &$\cal{P}$ \\ [0.5ex] \hline\hline
% Plus dagger
    \hspace{0.3cm} $\,A^{\dagger}_+ \bar{\lambda} P_+ \psi \, $ \hspace{0.3cm} & \hspace{0.3cm} $-\bar{\psi}  P_+ \lambda A_-^{\dagger}$ \hspace{0.3cm} & \hspace{0.3cm} $\, A_- \bar{\lambda} P_- \psi \,$  \hspace{0.3cm} \\[0.75ex]
    \hline
% Plus
    $\,\bar{\psi}  P_- \lambda A_+ $&$-A_- \bar{\lambda} P_- \psi \,$ & $ \,\bar{\psi}  P_+ \lambda A_-^{\dagger} \,$  \\[0.75ex]
    \hline
% Minus
    $\,A_- \bar{\lambda} P_- \psi  $&$-\bar{\psi}  P_- \lambda A_+ \,$ & $\, A^{\dagger}_+ \bar{\lambda} P_+ \psi \, $  \\[0.75ex]
    \hline 
% Minus dagger
    $\, \bar{\psi}  P_+ \lambda A_-^{\dagger}$&$-A^{\dagger}_+ \bar{\lambda} P_+ \psi \,$ & $ \,\bar{\psi}  P_- \lambda A_+ \, $  \\[0.75ex]
    \hline
%same as above but P_+ <-> P_-
$\,A^{\dagger}_+ \bar{\lambda} P_- \psi \, $&$-\bar{\psi}  P_- \lambda A_-^{\dagger}$ & $\, A_- \bar{\lambda} P_+ \psi \,$  \\[0.75ex]
    \hline
% Plus
    $\,\bar{\psi}  P_+ \lambda A_+ $&$-A_- \bar{\lambda} P_+ \psi \,$ & $ \,\bar{\psi}  P_- \lambda A_-^{\dagger} \,$  \\[0.75ex]
    \hline
% Minus
    $\,A_- \bar{\lambda} P_+ \psi  $&$-\bar{\psi}  P_+ \lambda A_+ \,$ & $\, A^{\dagger}_+ \bar{\lambda} P_- \psi \, $  \\[0.75ex]
    \hline 
% Minus dagger
    $\, \bar{\psi}  P_- \lambda A_-^{\dagger}$&$-A^{\dagger}_+ \bar{\lambda} P_- \psi \,$ & $ \,\bar{\psi}  P_+ \lambda A_+ \, $  \\[0.75ex]
    \hline

\hline
\end{tabular}
\caption{Gluino-squark-quark dimension-4 operators which are gauge invariant and flavor singlet. All matter fields carry an implicit flavor index.}
\label{tb:Ycoupling}
\end{center}
\end{table}

The transformation properties of the Yukawa terms, as shown in Table~\ref{tb:Ycoupling}, allow two distinct linear combinations of Yukawa-type operators: 
\bea
\label{chiInv}
&&Y_1 \equiv A^{\dagger}_+ \bar{\lambda} P_+ \psi   -  \bar{\psi}  P_- \lambda A_+ +  A_- \bar{\lambda} P_- \psi   -  \bar{\psi}  P_+ \lambda A_-^{\dagger} \\
&&Y_2 \equiv A^{\dagger}_+ \bar{\lambda} P_- \psi   -  \bar{\psi}  P_+ \lambda A_+ +  A_- \bar{\lambda} P_+ \psi   -  \bar{\psi}  P_- \lambda A_-^{\dagger}
\label{chiNonInv}
\eea
The first combination aligns with the third line of Eq.~(\ref{susylagrLattice}). However, at the quantum level, the second combination may emerge, having a potentially different Yukawa coupling. All terms within each of the combinations in Eqs.~(\ref{chiInv}) and~(\ref{chiNonInv}) are multiplied by a Yukawa coupling, denoted as $g_{Y_1}$ and $g_{Y_2}$ respectively. In the classical continuum limit, $g_{Y_1}$ corresponds to $g$, while $g_{Y_2}$ vanishes. 

Further symmetries to be considered are ${\cal{R}}$ and ${\cal{\chi}}$. The $U(1)_R$ symmetry, ${\cal{R}}$, rotates the quark and gluino fields in opposite direction:
\be
{\cal{R}}:\left \{\begin{array}{ll}
&\hspace{-.3cm} \psi(x)\rightarrow e^{i \theta \gamma_5}  \psi(x)\\[4pt]
&\hspace{-.3cm}\bar{ \psi}(x) \rightarrow\bar{ \psi}(x)e^{i \theta \gamma_5}\\[4pt]
&\hspace{-.3cm} \la(x)\rightarrow e^{-i \theta \gamma_5}  \la(x)\\[4pt]
&\hspace{-.3cm}\bar{ \la}(x) \rightarrow\bar{ \la}(x)e^{-i \theta \gamma_5}
\end{array}\right.
\label{Rsym}
\ee
${\cal{R}}$-symmetry does not commute with the SUSY transformation.
On the other hand the $U(1)_A$ symmetry, ${\cal{\chi}}$, rotates the squark and the quark fields in the same direction as follows: 
\be
{\cal{\chi}}:\left \{\begin{array}{ll}
&\hspace{-.3cm} \psi(x)\rightarrow e^{i \theta' \gamma_5}  \psi(x)\\[4pt]
&\hspace{-.3cm}\bar{ \psi}(x) \rightarrow\bar{ \psi}(x)e^{i \theta' \gamma_5}\\[4pt]
&\hspace{-.3cm} A_\pm(x) \rightarrow e^{ i \theta'} A_\pm(x)\\[4pt]
&\hspace{-.3cm} A_\pm^\dagger(x) \rightarrow e^{- i \theta'} A_\pm^\dagger(x)
\end{array}\right .
\label{chiral}
\ee
Eq.~(\ref{chiNonInv}) is not allowed as an extra Yukawa coupling counterterm on the lattice, if Ginsparg-Wilson gluinos are used. Both Yukawa terms commute with ${\cal R}$.  However the quark mass terms does not. Thus, if we insist on a theory with massive quarks, ${\cal{R}}$ is not a symmetry. ${\cal \chi} \times {\cal R}$ leaves invariant each of the four constituents of the Yukawa term (Eq.~(\ref{chiInv})), but it changes the constituents of the ``mirror'' Yukawa term (i.e. a term with all $P_+$ and $P_-$ interchanged) by phases $e^{2i \theta'}$ and $e^{-2i \theta'}$.

We note that the continuum action is invariant separately under ${\cal \chi}$ and ${\cal R}$ (for massless quarks), or under ${\cal \chi} \times {\cal R}$ (where the phases in ${\cal \chi}$ and ${\cal R}$ are chosen to be opposite, so that quarks are left unchanged) for massive quarks. The lattice action with Ginsparg-Wilson gluinos, even in the presence of Wilson quarks and/or a quark mass, will also be invariant under ${\cal \chi} \times {\cal R}$ (with opposite phases: $\theta = -\theta'$). In the same vein, if ${\cal \chi} \times {\cal R}$ is not anomalous,  there will be no mixing of  $A_+$ with $A_-^\dagger$ when using Ginsparg-Wilson gluinos, for the following reason: the Green's function e.g., $\langle A_+  A_- \rangle$, by a change of variables in the functional integral (corresponding to a ${\cal \chi} \times {\cal R}$  transformation of the integration variables) will become equal to $\langle A_+  A_- \rangle \times e^{2i \theta}$, and therefore it must vanish. But, indeed, in our paper~\cite{Costa:2017rht}, studying Wilson gluinos, we see that $\langle A_+  A_- \rangle$ does not vanish using the $HV$ definition of  $\gamma_5$. The same symmetry would guarantee also the absence of a ``mirror'' Yukawa term. Further, the mass term ``$M_2$'' in Ref.~\cite{Wellegehausen:2018opt}:
\be
M_2 \equiv A_+^\dagger  A_-^\dagger  +  A_-  A_+,
\ee
would also be absent in the presence of Ginsparg-Wilson gluinos (if ${\cal \chi} \times {\cal R}$ is not anomalous), since it is not invariant under ${\cal \chi} \times {\cal R}$. In our one-loop calculation of the lattice renormalization of the Yukawa couplings we have used Wilson gluinos, and indeed a counterterm of this type (Eq.~(\ref{chiNonInv})) has arisen.
% therefore, the critical value of masses are irrelevant since it is already of order $g^2$. 

\begin{figure}[ht!]
\centering
\includegraphics[scale=1.4]{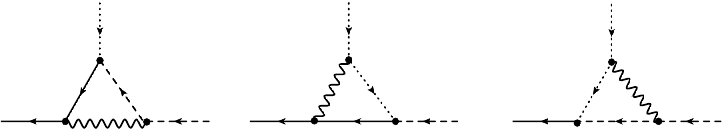}
\caption{One-loop Feynman diagrams leading to the fine tuning of $g_{Y_1}$ and $g_{Y_2}$. A wavy (solid) line represents gluons (quarks). A dotted (dashed) line corresponds to squarks (gluinos). %The ``double dashed'' line is the ghost field.
In the above diagrams the directions of the arrows on the external lines depend on the particular Green's function under study. An arrow entering (exiting) a vertex denotes a $\la, \psi, A_+, A_-^{\dagger}$ ($\bar \la, \bar \psi, A_+^{\dagger}, A_-$) field. Squark lines could be further marked with a $+$($-$) sign, to denote an $A_+ \, (A_-)$ field. 
}
\label{couplingYukawa}
\end{figure}

In our investigation, we compute perturbatively the relevant three-point Green's functions with external gluino, quark and squark fields, using both the $DR$ and the $LR$ regularizations. Each Green's function which contributes to the one-loop expression of the Yukawa couplings, consists of three Feynman diagrams shown in Fig.~\ref{couplingYukawa}. The renormalizations of fundamental fields and the gauge coupling are a prerequisite for the renormalization of the Yukawa coupling, since renormalization conditions in 3-pt-vertex corrections (with external gluino, quark and squark fields) involve these quantities. More specifically, combining the results for the bare Green's functions on the lattice with the renormalized Green's functions (obtained
in $\MSbar$ via DR), and using the renormalization factors for the gluino, quark, squark fields as well as the renormalization of the gauge coupling, we extract the renormalization and mixing factors of the Yukawa couplings appropriate to the lattice regularization and the $\MSbar$ renormalization scheme. 

Before we turn our attention to the calculation, notice that there exist several prescriptions~\cite{Larin:1993tq} for defining $\gamma_5$ in $D$ dimensions, such as the na\"ive dimensional regularization (NDR)~\cite{Chanowitz:1979zu}, the t'Hooft-Veltman (HV)~\cite{tHooft:1972}, the $DRED$~\cite{Siegel:1979} and the $DR{\overline{EZ}}$ prescriptions (see, e.g., Ref.~\cite{Patel}). They are related among themselves via finite conversion factors~\cite{Buras:1989xd}. In our calculation, we apply the NDR and HV prescriptions. The latter does not violate Ward identities involving pseudoscalar and axial-vector operators in $D$ dimensions~\cite{Chanowitz:1979zu}. The metric tensor, $\eta_{\mu\nu}$, and the Dirac matrices, $\gamma_\mu$, satisfy the following relations in $D$ dimensions:
\be
\eta^{\mu\nu}\eta_{\mu\nu}=d,\qquad \{\gamma_\mu,\gamma_\nu\} = 2 \eta_{\mu\nu} \openone.
\ee
In NDR, the definition of $\gamma_5$ satisfies:
\be
\{\gamma_5,\gamma_{\mu}\} = 0, \, \,\forall \mu,
\ee
whereas in HV it satisfies:
\be
\{\gamma_5,\gamma_{\mu}\} = 0, \, \,\mu = 1,2,3,4, \qquad [\gamma_5,\gamma_{\mu}]=0, \,\, \mu>4.
\ee

\section{Renormalization of the Yukawa couplings}
\label{couplingY}

In this section, we present our one-loop results for the bare 3-pt Green's functions and the renormalization factors of the Yukawa couplings in the $\MSbar$ scheme, using both dimensional ($DR$) and lattice ($LR$) regularizations. For the renormalization of $g_{Y_1}$ and $g_{Y_2}$, we impose renormalization conditions which result in the cancellation of divergences of the corresponding bare 3-pt amputated Green's functions with external gluino-squark-quark fields. The application of the renormalization factors on the bare Green's functions leads to the renormalized Green's functions, which are independent of the regulator ($\epsilon$ in $DR$, $a$ in $LR$). 

Given that we are interested in the $\MSbar$ renormalization of the Yukawa couplings, and that $\MSbar$ is a mass-independent renormalization scheme, we are free to treat all particles (in particular, quarks and squarks) as massless. In our forthcoming paper \cite{quartic:2023}, regarding the quartic (4-squark) couplings in SQCD, we choose instead to treat quarks and squarks as massive, in order to avoid the emergence of spurious infrared divergences. Such a scheme allows us to make use of techniques for evaluating Feynman diagrams which have been developed to very high perturbative order (see, e.g.,~\cite{Chetyrkin:1999pq, Ruijl:2017eht, Luthe:2017ttg, Chetyrkin:2017bjc, Gracey:2018fkg, Baikov:2019zmy}). Still perturbative calculations become exceedingly complicated on the lattice, and consequently, calculations beyond two loops are practically unfeasible.

The calculation of the amputated tree-level Green's functions is straight-forward and their expressions are\footnote{Note that the indices coming from the color in fundamental representation and the Dirac indices are left implicit. On the contrary, the color in the adjoint representation is shown explicitly.}:
\bea
%plusdagger in documentation (I have put the factor of 1/2 in alll GFs as discussed with Haris
\langle   \lambda^{\alpha_1}(q_1)  A_+(q_3) {\bar{\psi}} (q_2) \rangle^{\rm{tree}}  &=& - \frac{i}{2} \, g_{Y_1} \,(2\pi)^4 \delta(q_1-q_2+q_3) \, (1+\gamma_5) \, T^{\alpha_1}/ \sqrt 2\\
%plus in documentation
\langle  {\psi} (q_2)   A^{\dagger}_+(q_3)\bar \lambda^{\alpha_1}(q_1) \rangle^{\rm{tree}}  &=& \phantom{-}  \frac{i}{2} \, g_{Y_1} \, (2\pi)^4 \delta(q_1-q_2+q_3) \, (1-\gamma_5) \,T^{\alpha_1}/ \sqrt 2 \\
%minus in documentation
\langle   \lambda^{\alpha_1}(q_1)   A_-^{\dagger}(q_3)  {\bar{\psi}} (q_2)\rangle^{\rm{tree}}  &=& - \frac{i}{2} \, g_{Y_1} \,(2\pi)^4 \delta(q_1-q_2+q_3) \, (1-\gamma_5) \, T^{\alpha_1}/ \sqrt 2\\
%minusdagger in documentation
\langle  {\psi}(q_2)    A_-(q_3) \bar\lambda^{\alpha_1}(q_1) \rangle^{\rm{tree}}  &=& \phantom{-}  \frac{i}{2} \, g_{Y_1} \,(2\pi)^4 \delta(q_1-q_2+q_3) \,(1+\gamma_5) \, T^{\alpha_1}/ \sqrt 2
\eea
where our conventions for Fourier transformations are: 
\begin{align}
\tilde \psi(q) &= \int d^4x\, e^{-i q \cdot x}\,\psi(x), &  \psi(x) &= \int \frac{d^4q}{(2\pi)^4 } e^{i q \cdot x}\,\tilde \psi(q)\\
\tilde A_\pm(q) &= \int d^4x\, e^{\mp i q \cdot x}\,A_\pm(x), & A_\pm(x) &= \int \frac{d^4q}{(2\pi)^4 } e^{\pm i q \cdot x}\,\tilde A_\pm(q)\\
\tilde{u}_\mu(q) &= \int d^4x\, e^{-i q \cdot x}\,u_\mu(x), & u_\mu(x) &= \int \frac{d^4q}{(2\pi)^4 } e^{i q \cdot x}\,\tilde{u}_\mu(q)\\%
\tilde \la(q) &= \int d^4x\, e^{-i q \cdot x}\,\la(x), & \la(x) &= \int \frac{d^4q}{(2\pi)^4 } e^{i q \cdot x}\,\tilde \la(q)%\\
%\tilde c(q) &= \int d^4x\, e^{-i q \cdot x}\,c(x), & c(x) &= \int \frac{d^4q}{(2\pi)^4 } e^{i q \cdot x}\,\tilde c(q)%\\
\label{Fourier}
\end{align}  

The procedure of calculating the renormalization in the $\MSbar$ scheme entails performing first the perturbative calculations of the Green's function in $DR$; this is unavoidable by the very nature of the $\MSbar$ scheme. The comparison with the same Green's functions calculated in $LR$ will lead to the lattice renormalizations in the $\MSbar$ scheme. 

The calculations presented in this paper could ideally be performed using generic external momenta. However, for convenience of computation, we are free to make appropriate choices of these momenta; the resulting renormalization factors will not be affected at all. Having checked that no superficial $IR$ divergences will be generated, we calculate the corresponding diagrams by setting to zero only one of the three external momenta. The choice of the external momenta for Green's functions will not affect their pole parts in $DR$ or their logarithmic dependence on the lattice spacing in $LR$. Therefore, there are three choices for each 3-pt Green's function depending on which momentum is set zero. 

There are, in total, 4 different gluino-squark-quark Green's functions, depending on whether the external squark field is $A_+ / A_+^{\dagger} / A_- / A_-^{\dagger}$. We present first the four Green's functions for the three choices of external momentum in $DR$. To avoid heavy notation we have omitted Dirac/flavor/color indices\footnote{The color indices in the adjoint representation are shown explicitly.} on the Green's functions of Eqs.~(\ref{plusdagger})-(\ref{SMom0}). 
\bea
\label{plusdagger}
%Plus dagger
\langle   \lambda^{\alpha_1}(0)  A_+(q_3) {\bar{\psi}} (q_2)  \rangle^{DR, {\rm{1loop}}}  &=& - \, \langle  {\psi}(q_2)    A_-(q_3) \bar\lambda^{\alpha_1}(0) \rangle^{DR, {\rm{1loop}}} = -i\,(2\pi)^4 \delta(q_2-q_3) \frac{g_{Y_1} g^2}{16\pi^2} \frac{1}{4\sqrt2 N_c} T^{\alpha_1} \times \nonumber\\
&&\hspace{-3.75cm}  \Bigg[-3 (1 + \gamma_5) + ((1 + \alpha)(1 + \gamma_5) + 8 \gamma_5 c_{\rm hv}) N_c^2  + (1 + \gamma_5) (-\alpha + (3 + 2 \alpha) N_c^2 ) \left(\frac{1}{\epsilon}+ \log\left(\frac{\bar \mu^2}{q_2^2}\right) \right)\Bigg]%gMom0  
\\
\langle   \lambda^{\alpha_1}(q_1) A_+(q_3) {\bar{\psi}} (0)  \rangle^{DR, {\rm{1loop}}}  &=& - \, \langle  {\psi}(0)    A_-(q_3) \bar\lambda^{\alpha_1}(q_1) \rangle^{DR, {\rm{1loop}}} = -i\,(2\pi)^4 \delta(q_1+q_3) \frac{g_{Y_1} g^2}{16\pi^2} \frac{1}{4\sqrt2 N_c} T^{\alpha_1} \times \nonumber\\ 
&&\hspace{-3.75cm} \Bigg[ ( (4 + \alpha) (1 + \gamma_5) + 8 \gamma_5 c_{\rm hv}) N_c^2 + (1 + \gamma_5) (-\alpha + (3 + 2 \alpha) N_c^2 ) \left(\frac{1}{\epsilon}+ \log\left(\frac{\bar \mu^2}{q_1^2}\right) \right)\Bigg]%QMom0  
\\
\label{Plus}
\langle   \lambda^{\alpha_1}(q_1) A_+(0) {\bar{\psi}} (q_2)  \rangle^{DR, {\rm{1loop}}}  &=& - \, \langle  {\psi}(q_2)    A_-(0) \bar\lambda^{\alpha_1}(q_1) \rangle^{DR, {\rm{1loop}}} = - i \,(2\pi)^4 \delta(q_1-q_2) \frac{g_{Y_1} g^2}{16\pi^2} \frac{1}{4\sqrt2 N_c} T^{\alpha_1} \times \nonumber\\ 
&&\hspace{-3.75cm} \Bigg[-\alpha (1 + \gamma_5) + ((4 + 3 \alpha) (1 + \gamma_5) + 8 \gamma_5 c_{\rm hv}) N_c^2 + (1 + \gamma_5) (-\alpha + (3 + 2 \alpha) N_c^2 ) \left(\frac{1}{\epsilon}+ \log\left(\frac{\bar \mu^2}{q_1^2}\right) \right)\Bigg]%SMom0  
\eea
\bea
%Plus
\langle  {\psi} (q_2)   A^{\dagger}_+(q_3)\bar \lambda^{\alpha_1}(0) \rangle^{DR, {\rm{1loop}}}  &=& - \, \langle   \lambda^{\alpha_1}(0)   A_-^{\dagger}(q_3)  {\bar{\psi}} (q_2)\rangle^{DR, {\rm{1loop}}} = - i \,(2\pi)^4 \delta(q_2-q_3) \frac{g_{Y_1} g^2}{16\pi^2} \frac{1}{4\sqrt2 N_c} T^{\alpha_1} \times  \nonumber\\ 
&&\hspace{-3.75cm}\Bigg[ 3(1-\gamma_5) - ((1 + \alpha) (1-\gamma_5) - 8\gamma_5 c_{\rm hv}) N_c^2 - (1 - \gamma_5) (-\alpha + (3 + 2 \alpha) N_c^2 ) \left(\frac{1}{\epsilon}+ \log\left(\frac{\bar \mu^2}{q_2^2}\right) \right)\Bigg]%gMom0  
\\
\langle  {\psi} (0)   A^{\dagger}_+(q_3)\bar \lambda^{\alpha_1}(q_1)  \rangle^{DR, {\rm{1loop}}}  &=&  -\, \langle   \lambda^{\alpha_1}(q_1) A_-^{\dagger}(q_3) {\bar{\psi}} (0)  \rangle^{DR, {\rm{1loop}}} =  - i\,(2\pi)^4 \delta(q_1+q_3) \frac{g_{Y_1} g^2}{16\pi^2} \frac{1}{4\sqrt2 N_c} T^{\alpha_1} \times \nonumber\\ 
&&\hspace{-3.75cm}\Bigg[(-(4 + \alpha) (1-\gamma_5) N_c^2 +8 \gamma_5 c_{\rm hv}) N_c^2 - (1 - \gamma_5) (-\alpha + (3 + 2 \alpha) N_c^2 ) \left(\frac{1}{\epsilon}+ \log\left(\frac{\bar \mu^2}{q_1^2}\right) \right)\Bigg]%QMom0
\\
\langle  {\psi} (q_2)   A^{\dagger}_+(0)\bar \lambda^{\alpha_1}(q_1) \rangle^{DR, {\rm{1loop}}}  &=& - \, \langle   \lambda^{\alpha_1}(q_1) A_-^{\dagger}(0)  {\bar{\psi}} (q_2) \rangle^{DR, {\rm{1loop}}} =-i\,(2\pi)^4 \delta(q_1-q_2) \frac{g_{Y_1} g^2}{16\pi^2} \frac{1}{4\sqrt2 N_c} T^{\alpha_1} \times\nonumber\\ 
&&\hspace{-3.75cm} \Bigg[ \alpha (1- \gamma_5) + (-(4 + 3 \alpha) (1 - \gamma_5) + 8 \gamma_5 c_{\rm hv}) N_c^2- (1 - \gamma_5) (-\alpha + (3 + 2 \alpha) N_c^2 ) \left(\frac{1}{\epsilon}+ \log\left(\frac{\bar \mu^2}{q_1^2}\right) \right)\Bigg]%SMom0
\label{SMom0}
\eea
where  $c_{\rm hv} = 0 \, (1)$ for the NDR (HV) prescription \cite{Costa:2017rht} of $\gamma_5$. The pole parts do not depend on $c_{\rm hv}$. Further, in the NDR prescription, all one-loop bare Green's functions are proportional to the tree-level ones. The above one-loop Green's functions indeed confirm that the pole parts are the same for different choices of the external momenta and that they are proportional to the tree-level value. In $HV$, due to the fact that the first quantum corrections (one-loop) of these Green's functions have finite parts which are not proportional to their tree-level counterparts (i.e., in addition to terms with ($1\pm\gamma_5$), they contain also terms with ($1\mp\gamma_5$)), it seems that the chiral symmetry is anomalous even in $DR$. The need of further conversion factors, which connect $\MSbar$ renormalized Green's functions to SUSY invariant Green's functions, is indicated by the supersymmetric Ward Identities~\cite{Wellegehausen:2018opt}. The value of these conversion factors requires a purely continuum calculation, including Eqs.~(\ref{plusdagger})-(\ref{SMom0}); the same conversion factors can be applied to the renormalization functions extracted in $LR$.  

Note that the terms in Eqs.~(\ref{plusdagger})-(\ref{SMom0}) involving multiplication by $c_{\rm hv} \gamma_5$ can be equivalently expressed as: $\frac{1}{2} c_{\rm hv} \left( (1+\gamma_5) - (1-\gamma_5) \right)$. Terms with reversed chirality account for the mirror Yukawa interactions; given that they are pole free and they will have no effect on a straightforward $\MSbar$ renormalization. However, if one opts for a renormalization scheme in which these terms are absent, one must add a finite $Y_2$ counterterm to the action of the form: 
\be
{\cal L}^{\rm ct}_{Y_2} \equiv i \, \sqrt{2} \, g_{Y_2} Y_2, \, \, \,  {\rm where: \, \, } g_{Y_2} = 2 g^3 N_c \, c_{\rm hv} / (16\pi^2) + {\cal{O}}(g^5)  .
\label{counterterm}
\ee
This term, as well as Eqs.~(\ref{plusdagger})-(\ref{SMom0}), become relevant in our lattice calculations as they contribute to finite mixing coefficients. %It is worth noting that this mixing coefficient is absent in $DR$ and the $\MSbar$ scheme due to the pole-free nature of the mirror Yukawa terms at the one-loop order.

The difference between the renormalized Green's functions and the corresponding Green's functions regularized on the lattice allows us to deduce the one-loop lattice renormalizations factors.  The renormalization factors of the fields and the gauge coupling constant can be found in Ref.~\cite{Costa:2017rht}. For the sake of completeness we present their definition here:
\bea
\psi \equiv \psi^B &=& Z_\psi^{-1/2}\, \psi^R,\\
%A^R_\pm &=& \sqrt{Z_{A_\pm}}\,A^B_\pm, \\
u_{\mu} \equiv u_{\mu}^B &=& Z_u^{-1/2}\,u^R_{\mu},\\
\la \equiv \la^B &=& Z_\la^{-1/2}\,\la^R,\\
c \equiv c^B &=& Z_c^{-1/2}\,c^R, \\
g \equiv g^B &=& Z_g^{-1}\,\mu^{\epsilon}\,g^R \label{g}, 
\eea
where $B$ stands for the bare and $R$ for renormalized quantities and $\mu$ is an arbitrary scale with dimensions of inverse length. For one-loop calculations, the distinction between $g^R$ and $g^B$ is inessential in many cases; we will simply use $g$ in those cases. 
The Yukawa coupling is renormalized as follows: 
\be
\label{gy}
g_{Y_1} \equiv g_{Y_1}^B = Z_{Y_1}^{-1} Z_g^{-1} \mu^\epsilon g^R,
\ee
where at the lowest perturbative order $Z_g Z_{Y_1} = 1$, and the renormalized Yukawa coupling coincides with the gauge coupling. 

In $DR$, we are interested in getting rid of the pole parts in bare continuum Green's functions; this requires not only the renormalization factors of the fields and of the gauge coupling, $Z_g$, but also a further factor $Z_{Y_1}$ for the bare Yukawa coupling. Note also that the components of the squark fields may mix at the quantum level, via a $2\times2$ mixing matrix ($Z_A$). We define the renormalization mixing matrix for the squark fields as follows:
\be
\label{condS}
\left( {\begin{array}{c} A^R_+ \\ A^{R\,\dagger}_- \end{array} } \right)= \left(Z_A^{1/2}\right)\left( {\begin{array}{c} A^B_+ \\ A^{B\,\dagger}_- \end{array} } \right).
\ee
In Ref.~\cite{Costa:2017rht} we found that in the $DR$ and $\MSbar$ scheme this $2\times2$ mixing matrix is diagonal. On the lattice, this matrix is non-diagonal, leading to a mixing of the components $A_+$ and $A_-$ with $A_-^\dagger$ and $A_+^\dagger$, respectively. Consequently, the renormalization conditions on the lattice become more intricate. In this paper we focus on the $\MSbar$ scheme, using both $DR$ and $LR$ regularizations. Given that SUSY is broken by either regulator and that SUSY-noninvariant gauge fixing is employed, it is anticipated that a nontrivial fine-tuning for the Yukawa coupling will be necessary.

Taking as an example the Green's function in $DR$ with external squark field $A_+$, the renormalization condition up to $g^2$ will be given by:
\be
\langle   \lambda(q_1)  A_+(q_3) {\bar{\psi}} (q_2) \rangle \Big \vert^\MSbar = Z_\psi^{-1/2} Z_\la^{-1/2}  (Z_A^{-1/2})_{++} \langle   \lambda(q_1)  A_+(q_3) {\bar{\psi}} (q_2) \rangle \Big \vert^{\rm{bare}} 
\label{renormC}
\ee
All appearances of coupling constants in the right-hand side of Eq.~(\ref{renormC}) must be expressed in terms of their renormalized values, via Eqs.~(\ref{g})-(\ref{gy}). The left-hand side of Eq.~(\ref{renormC}) is just the $\MSbar$ (free of pole parts) renormalized Green's function. Similar to Eq.~(\ref{renormC}), the other renormalization conditions which involve the external squark fields $A_+^\dagger, A_-, A_-^\dagger$ are understood. The renormalization factors $Z = \openone + {\cal O}(g^2)\,$ and mixing coefficients $z = {\cal O}(g^2)\,$ should more properly be denoted as $Z^{X,Y}$ and $z^{X,Y}$, where $X$ is the regularization and $Y$ the renormalization scheme. 

For the sake of clarity and comprehensiveness, the updated expressions for the renormalization factors of the fields and of the gauge coupling in $DR$ which are involved in the right-hand side of Eq.~(\ref{renormC}) are
\footnote{The expressions for $Z_{\psi}, Z_{A_\pm}, Z_{\lambda}$ and $Z_g$ (Eqs.~(\ref{zp})-(\ref{zg}), Eqs.~(\ref{zpl})-(\ref{zgl})) appeared also in Ref.~\cite{Costa:2017rht}; however, a factor of 1/2 was missing in diagrams involving open internal gluino lines. For a more detailed explanation, see Appendix \ref{majoranaS}.}:
\bea
\label{zp}
Z_\psi^{DR,\MSbar} &=& 1 + \frac{g^2\,C_F}{16\,\pi^2} \frac{1}{\epsilon}\left( 1 + \alpha\right)\\
Z_{A_\pm}^{DR,\MSbar} &=& 1 + \frac{g^2\,C_F}{16\,\pi^2} \frac{1}{\epsilon}\left(-1 + \alpha \right)\\
Z_{\lambda}^{DR,\MSbar} &=&  1 + \frac{g^2\,}{16\,\pi^2} \frac{1}{\epsilon} \left(\alpha\, N_c + N_f \right)\\
Z_{g}^{DR,\MSbar}&=&1 + \frac{g^2\,}{16\,\pi^2} \frac{1}{\epsilon} \left(\frac{3}{2} N_c - \frac{1}{2}N_f \right),
\label{zg}
\eea
where $C_F=(N_c^2-1)/(2\,N_c)$ is the quadratic Casimir operator in the fundamental representation. The expressions in Eqs.~(\ref{zp})-(\ref{zg}) take carefully into account the effect of the Majorana nature of gluinos in the functional integral. In Appendix \ref{majoranaS}, we provide a more comprehensive discussion and treatment of the gluino field; in particular, we focus on the effect of Yukawa terms in SQCD, which are clearly absent in pure SUSY Yang-Mills. 

Substituting Eqs.~(\ref{zp})-(\ref{zg}) in Eq.~(\ref{renormC}), and by virtue of the fact that the counterterm Eq.~(\ref{counterterm}) contains no pole parts, we extract the value of $Z_{Y_1}^{DR, \MSbar}$; this value is the same for all gluino-squark-quark Green's functions and for all choices of the external momenta which we have considered:
\be
Z_{Y_1}^{DR, \MSbar}= 1 + {\cal O}(g^4)  \quad
\label{ZYDR}
\ee
%where $C_F=(N_c^2-1)/(2\,N_c)$ is the quadratic Casimir operator in the fundamental representation. As expected from general renormalization theorems, the $\MSbar$ renormalization factors for gauge invariant objects are gauge-independent as in the case of $Z_Y^{DR, \MSbar}$.
Eq.~(\ref{ZYDR}) means that, at the quantum-level, the renormalization of the Yukawa coupling in DR is not affected by one-loop corrections. This observation has important implications for our understanding of the renormalization scheme in SQCD. It shows also that the corresponding renormalization on the lattice will be finite. Although, the mirror Yukawa term does not appear in $\MSbar$ renormalization using DR, a finite mixing with this term will arise in $\MSbar$ on the lattice. We expect that the $\MSbar$ renormalization factors of gauge invariant quantities will turn out to be gauge-independent also on the lattice, as was the case of $Z_{Y_1}^{DR, \MSbar}$.

\vspace{1.2 cm}
 
We now turn to the lattice regularization.  As emphasized earlier, even though the renormalization of the squark fields in the $\MSbar$ scheme and in $DR$ is diagonal, on the lattice it is not; the mixing between the squark components ($A_+$,$A_-^{\dagger}$) (and, similarly, ($A_+^{\dagger}$, $A_-$)) appears on the lattice through the $2\times2$ symmetric matrix $Z_A$, whose nondiagonal matrix elements are nonzero. The renormalization conditions are not as simple as is shown in Eq.~(\ref{renormC}); instead, they involve the following pairs of Green's functions:
\bea
\langle   \lambda(q_1)  A_+(q_3) {\bar{\psi}} (q_2)  \rangle  &{\rm with}& \langle   \lambda(q_1) A_-^{\dagger}(q_3)  {\bar{\psi}}   (q_2) \rangle  \nonumber \\ 
\langle  {\psi} (q_2)   A^{\dagger}_+(q_3)\bar \lambda(q_1) \rangle  &{\rm with} &  \langle  {\psi}(q_2)    A_-(q_3) \bar\lambda(q_1)  \rangle
\eea
The appearance of the mirror Yukawa coupling, $g_{Y_2}$, is another feature of the use of Wilson gluinos, which increases the degree of difficulty on the lattice. The ${\cal \chi} \times {\cal R}$ symmetry is broken by using Wilson discretization and thus lattice bare Green's functions are not invariant under ${\cal \chi} \times {\cal R}$ at the quantum level. This difficulty may be avoided with chirality preserving actions, but the implementation of these actions in numerical simulations is very time consuming. 

%It is convenient to group the Yukawa couplings as:
%\be
%g_Y =\left( {\begin{array}{c} g_{Y_1} \\ g_{Y_2} \end{array} } \right)
%\ee
%Then the renormalization of the Yukawa coupling on the lattice is given by a $2\times2$ matrix.  The first matrix element of the renormalization factor of the Yukawa coupling on the lattice is chosen in such a way so that:
%where overall powers of the lattice spacing are omitted. 

Thus, in the calculation of bare Green's functions on the lattice, one-loop spurious contributions will arise, which will need to be removed by introducing mirror Yukawa counterterms in the action. The renormalization condition is the following: 
\be
\langle   \lambda(q_1)  A_+(q_3) {\bar{\psi}} (q_2) \rangle \Big \vert^\MSbar = Z_\psi^{-1/2} Z_\la^{-1/2}  \langle   \lambda(q_1)  \bigl((Z_A^{-1/2})_{++} A_+(q_3) + (Z_A^{-1/2})_{+-} A_-^\dagger(q_3)\bigr){\bar{\psi}} (q_2)  \rangle \Big \vert^{\rm{bare}} 
% \nonumber \\\langle    {\psi} (q_2)   A^{\dagger}_a(q_3)\bar \lambda(q_1)\rangle^{\rm{tree\, + \,1loop}}_{\rm{amp}}\Big \vert^\MSbar &=& Z_\psi^{-1/2} Z_\la^{-1/2}  (Z_A^{-1/2})_{a b} Z_g^{-1} (Z_Y^{-1})^{{b c}} \langle {\psi} (q_2)   A^{\dagger}_c(q_3)\bar \lambda(q_1) \rangle^{\rm{tree \,+\, 1loop}}_{\rm{amp}}\Big \vert^{\rm{bare}}
\label{renormCLatt}
\ee
It is understood that the bare couplings on the right-hand side of this equation must be converted into the corresponding renormalized ones, making use of $Z_g$ and $Z_{Y_1}$; a mirror Yukawa term also contributes, with a coupling constant $g_{Y_2}$ which will be determined in what follows. Eq. (\ref{renormCLatt}) consists of two types of contributions with opposite chiralities; matching each of these to the $\MSbar$ expressions found in DR, Eqs.~(\ref{plusdagger})-(\ref{Plus}), amounts
to two separate conditions, which will be used to determine the two unknowns $Z_{Y_1}$ and $g_{Y_2}$. Analogous equations hold for the other gluino-squark-quark Green's functions and may be calculated for consistency checks.

To offer a self-contained presentation, we revisit a collection of lattice results outlined in Ref.~\cite{Costa:2017rht}. 
\bea
\label{zpl}
Z_\psi^{LR,\MSbar} &=& 1 + \frac{g^2\,C_F}{16\,\pi^2} \left( -16.7235 + 3.7920 \alpha - (1+\alpha)\log\left(a^2\,\bar\mu^2\right) \right)\\
\left(Z_A^{1/2}\right)^{LR,\MSbar} &=&  \openone - \,\frac{g^2\,C_F}{16\,\pi^2}\Bigg\{\Bigg[16.9216 - 3.7920\alpha-(1-\alpha)\log\left(a^2\,\bar\mu^2\right)\Bigg] \begin{pmatrix} 1 & 0\\ 0 & 1 \end{pmatrix} - 0.1623 \begin{pmatrix} 0 & 1\\ 1 & 0 \end{pmatrix}\Bigg\}\\
Z_{\lambda}^{LR,\MSbar} &=&  1 - \frac{g^2\,}{16\,\pi^2} \left[N_c\left(16.6444 - 3.7920 \alpha + 2\, \alpha \log\left(a^2\,\bar\mu^2\right)\right)+ N_f\left(0.07907 + 2\log\left(a^2\,\bar\mu^2\right)\right) \right] \\
Z_g^{LR,\MSbar} &=&  1 + \gtilde\,\Bigg[ -9.8696 \frac{1}{N_c} + N_c \left( 12.8904  - \frac{3}{2} \log\left(a^2\,\bar{\mu}^2\right)\right)-\,N_f\left( 0.4811 - \frac{1}{2} \log(a^2\,\bar{\mu}^2)\right)\Bigg]
\label{zgl}
\eea

The lattice 3-pt Green's functions involve the same Feynman diagrams as in Fig.~\ref{couplingYukawa}. At first perturbative order, ${\cal O}(g^2)\,$, Eq.~(\ref{renormCLatt}) and its counterparts involve only the difference between the one-loop $\MSbar$-renormalized and bare lattice Green's functions. Having checked that alternative choices of the external momenta give the same results for these differences, we present them only for zero gluino momentum. Additionally, we should mention that the errors on our lattice expressions are smaller than the last shown digit and the Wilson parameter, $r$ was set to its default value: $r=1$.
\bea
%Plus dagger
&&\hspace{-1cm} \langle   \lambda^{\alpha_1}(0)  A_+(q_3) {\bar{\psi}} (q_2)  \rangle^{\MSbar, {\rm{1loop}}}  - \langle   \lambda^{\alpha_1}(0)  A_+(q_3) {\bar{\psi}} (q_2)  \rangle^{LR, {\rm{1loop}}}  = - \langle  {\psi}(q_2)    A_-(q_3) \bar\lambda^{\alpha_1}(0) \rangle^{\MSbar, {\rm{1loop}}}  + \langle  {\psi}(q_2)    A_-(q_3) \bar\lambda^{\alpha_1}(0) \rangle^{LR, {\rm{1loop}}} \nonumber\\
&&\hspace{-0.25cm} =i\,(2\pi)^4 \delta(q_2-q_3) \frac{g_{Y_1} g^2}{16\pi^2} \frac{1}{8\sqrt2 N_c} T^{\alpha_1}\times \Bigg[-3.7920 \alpha ( 1 + \gamma_5) + (-3.6920 + 5.9510 \gamma_5 +7.5840 \alpha ( 1 + \gamma_5) -  8 \gamma_5 c_{\rm hv}) N_c^2 
\nonumber\\ 
&&\hspace{5.97cm} + (1 + \gamma_5) (\alpha - (3 + 2 \alpha) N_c^2 ) \log\left(a^2 \bar \mu^2 \right) 
\Bigg]%gMom0  
\\
%\langle   \lambda^{\alpha_1}(q_1)  {\bar{\psi}} (0) A_+(q_3) \rangle^{\MSbar, {\rm{1loop}}}  - \langle   \lambda^{\alpha_1}(q_1)  {\bar{\psi}} (0) A_+(q_3) \rangle^{LR, {\rm{1loop}}}  &=& -i\,(2\pi)^4 \delta(q_1+q_3) \frac{g_Y g^2}{16\pi^2} \frac{1}{4\sqrt2 N_c} T^{\alpha_1}\times \nonumber\\ 
%&&\hspace{-10.75cm} \Bigg[ -3.7920 \alpha ( 1 + \gamma_5) + (-3.6920 + 5.9510 \gamma_5 +7.5840 \alpha ( 1 + \gamma_5) -  8 \gamma_5 c_{\rm hv}) N_c^2 + (1 + \gamma_5) (\alpha - (3 + 2 \alpha) N_c^2 ) \log\left(a^2 \bar \mu^2 \right)
%\Bigg]%QMom0  
%\\
%\langle   \lambda^{\alpha_1}(q_1)  {\bar{\psi}} (q_2) A_+(0) \rangle^{\MSbar, {\rm{1loop}}}  - \langle   \lambda^{\alpha_1}(q_1)  {\bar{\psi}} (q_2) A_+(0) \rangle^{LR, {\rm{1loop}}}  &=& i\,(2\pi)^4 \delta(q_1-q_2) \frac{g_Y g^2}{16\pi^2} \frac{1}{4\sqrt2 N_c} T^{\alpha_1} \times \nonumber\\ 
%&&\hspace{-10.75cm}\Bigg[-3.7920 \alpha ( 1 + \gamma_5) + (-3.6920 + 5.9510 \gamma_5 +7.5840 \alpha ( 1 + \gamma_5) -  8 \gamma_5 c_{\rm hv}) N_c^2 + (1 + \gamma_5) (\alpha - (3 + 2 \alpha) N_c^2 ) \log\left(a^2 \bar \mu^2 \right)
%\Bigg]%SMom0  
%\eea
%\bea
%Plus
&&\hspace{-1cm}\langle  {\psi} (q_2)   A^{\dagger}_+(q_3)\bar \lambda^{\alpha_1}(0) \rangle^{\MSbar, {\rm{1loop}}}  - \langle  {\psi} (q_2)   A^{\dagger}_+(q_3)\bar \lambda^{\alpha_1}(0) \rangle^{LR, {\rm{1loop}}} = -\langle   \lambda^{\alpha_1}(0)   A_-^{\dagger}(q_3)  {\bar{\psi}} (q_2)\rangle^{\MSbar, {\rm{1loop}}}  + \langle   \lambda^{\alpha_1}(0)   A_-^{\dagger}(q_3)  {\bar{\psi}} (q_2)\rangle^{LR, {\rm{1loop}}} \nonumber \\
&&\hspace{-0.25cm} = i\,(2\pi)^4 \delta(q_2-q_3) \frac{g_{Y_1} g^2}{16\pi^2} \frac{1}{8\sqrt2 N_c} T^{\alpha_1}\times \Bigg[
 3.79201 \alpha ( 1 - \gamma_5) + (3.6920 + 5.9510 \gamma_5 -7.5840 \alpha ( 1- \gamma_5) -  8 \gamma_5 c_{\rm hv}) N_c^2 
 \nonumber\\ 
&&\hspace{5.97cm} +(1 - \gamma_5) (-\alpha + (3 + 2 \alpha) N_c^2 ) \log\left(a^2 \bar \mu^2 \right)
\Bigg]%gMom0  
\\
 \nonumber
\eea
As expected, the above expressions are momentum-independent, and they are linear combinations of the tree-level expressions stemming from the Yukawa vertex and its mirror; also, all corresponding decimal coefficients between Eqs. (69) and (70) coincide, and we have checked that they are the same for any other choice of external momenta, as they should. Thus, we are led to a unique result for ${Z_Y}^{LR,\MSbar}$ and also for ${g_{Y_2}}^{LR,\MSbar}$. By combining the lattice expressions with the $\MSbar$-renormalized Green's functions calculated in the continuum (see Eq.~(\ref{renormCLatt})), we find for the renormalization factors: 
\bea
{Z_{Y_1}}^{LR,\MSbar} &=& 1 + \frac{g^2}{16\,\pi^2}  \left(\frac{1.45833}{N_c} + (4.40768- 2 c_{\rm hv})N_c + 0.520616 N_f \right)\\
{g_{Y_2}}^{LR,\MSbar} &=&\frac{g^3}{16\,\pi^2}\left(\frac{-0.040580}{N_c} + (2.45134 - 2 c_{\rm hv} )N_c \right)
\eea
We note that the above factors are gauge independent in the $\MSbar$ scheme, as expected from the principles of renormalization and gauge invariance. Furthermore, the multiplicative renormalization $Z_{Y_1}^{LR,\MSbar}$ and the coefficient $g_{Y_2}^{LR,\MSbar}$ of the mirror Yukawa counterterm are finite as one can predict from the continuum calculation. These findings shed light on the fine-tunings for the lattice SQCD action. They suggest that while the renormalization process in $\MSbar$ is well-behaved on the lattice, it still exhibits an intriguing connection with the mirror Yukawa term through $g_{Y_2}^{LR,\MSbar}$.

\section{Outlook -- Future Plans}
\label{summary}

In this work we calculate 3-pt Green's functions with external elementary fields for the SQCD action in the Wess-Zumino gauge. In particular, we perform one-loop calculations for a complete set of 3-pt Green's functions with external gluino, quark and squark fields, employing Wilson fermions and gluons. To extract the fine-tunings of Yukawa couplings in the $\MSbar$ scheme, we compute the relevant Green's functions in two regularizations: dimensional and lattice. The lattice calculations are the crux of this work; and the continuum calculations serve as a necessary ingredient, allowing us to relate our lattice results to the $\MSbar$ scheme.

 With the perturbative renormalization of the Yukawa couplings we make a step forward on the completion of all renormalizations (fields, masses, couplings) in the Wilson formulation~\cite{Costa:2017rht, Costa:2018mvb}. The results of this work will be particularly relevant for the setup and the calibration of lattice numerical simulations of SQCD. In the coming years, it is expected that simulations of supersymmetric theories will become ever more feasible and precise. 
 
 A follow up calculation regards the quartic couplings (four-squark interactions). The symmetries allow five quartic couplings~\cite{Wellegehausen:2018opt}, which must be also appropriately fine tuned on the lattice.  This is a natural extension of our work and the calculation of their quantum corrections is currently underway~\cite{Herodotou:2022xhz}.
 
\appendix
\section{The path integral over the gluino field}
\label{majoranaS}
To elucidate the Majorana nature of the gluino field within the functional integral, and the way to properly address it in the calculation of Feynman diagrams, we first reformulate the action from Eq.~(\ref{susylagr}) to express it in exclusively in terms of $\lambda$, rather than $\bar \lambda$. We proceed in a way analogous to Ref.~\cite{Donini:1997hh}, but we now take into account the additional complication brought about by the Yukawa terms. By applying the Majorana condition ($(\bar \lambda^{\alpha})^T= C\lambda^{\alpha}$), the part of the action which contains gluino fields has the general form:
\begin{equation}
S_{\text{gluino}}= \bar \lambda D \lambda + \bar{A} \lambda + \bar{\lambda} B = \lambda^T M \lambda + (\bar{A} + B') \lambda,
\end{equation}
where $M \equiv C D$. The first term represents both the kinetic energy of the gluino and the interaction with the gluon field. The subsequent terms correspond to the Yukawa interactions:
\begin{equation}
\bar{A}= i \sqrt{2} \, g \, (- \bar{\psi} P_- T^{\alpha} A_+ - \bar{\psi} P_+ T^{\alpha} A_-^{\dagger}), \quad B= i \sqrt2 \, g \, ( A^{\dagger}_+  T^{\alpha} P_+ \psi +  A_-  T^{\alpha} P_- \psi )
\end{equation}
where $B'= -B^T C$ and $B'^{T}= C B$. Therefore, the path integral reads:
\begin{equation}
    Z[J] = \int \mathcal{D} U_{\text{other}} \,  e^{-S_{\text{other}}} \int \mathcal{D} \lambda \, e^{- \lambda^T M \lambda - (\bar{A} + B') \lambda - J \lambda } \, ,
    \label{PartFun}
\end{equation}
where $J$ is an external source, $U_{\text{other}}$ stands for all of the fields in the theory except gluino fields, and $S_{\text{other}}$ denotes the action part devoid of gluinos. In order to integrate out the gluino field, we implement the following standard change of variables:
\begin{equation}
    \la'^T \equiv \la^T + \frac{1}{2} ( J + \bar{A} + B' ) \, M
\end{equation}
This leads to:
\begin{align}
    Z[J] &= \int \mathcal{D} U_{\text{other}} \,  e^{-S_{\text{other}}} \int \mathcal{D} \lambda' \, e^{- \lambda'^T M \lambda' - \frac{1}{4} (\bar{A} + B' + J) M^{-1} (\bar{A} + B' + J)^T} \nonumber \\
    &= \int \mathcal{D} U_{\text{other}} \, e^{-S_{\text{other}}} \, Pf[M]  \, e^{- \frac{1}{4} (\bar{A} + B' + J) M^{-1} (\bar{A} + B' + J)^T}
    \label{ZJeq}
\end{align}
where $Pf[M]$ is the Pfaffian of the antisymmetric matrix $M$. In the absence of Yukawa terms, and in case one is interested only in Green's functions without external gluinos (so that one can set $J=0$ from the start), the exponential in Eq.~(\ref{ZJeq}) becomes trivial and the only remnant of gluinos is the Pfaffian; in those cases, the only effect of the gluinos' Majorana nature is the well-known factor of 1/2 for every closed gluino loop, due to the fact that $Pf[M] = \rm{det}[M]^{1/2}$. Note that we do not assume that $J$, $\bar{A}$ and $B$ are Majorana spinors. Let us examine the exponent appearing in Eq.~(\ref{ZJeq}):
\begin{align}
    -S' \equiv -\frac{1}{4} ( \bar{A} + B' + J) M^{-1} (\bar{A} + B' + J)^T
\label{Seff}
\end{align}

When we compute Green's functions without external gluinos, we can set $J=0$ and thus, $S'$ can be written as:
\begin{align}
    -S'|_{J=0}&= -\frac{1}{4} ( \bar{A} + B') M^{-1} (\bar{A} + B')^T \nonumber \\
    &= -\frac{1}{4} \big( \bar{A} M^{-1} \bar{A}^T + B' M^{-1} B'^T + \bar{A} M^{-1} C B - B^T C M^{-1} \bar{A}^T \big) \nonumber \\
    &= -\frac{1}{4} \big( \bar{A} M^{-1} \bar{A}^T + B' M^{-1} B'^T + 2 \, \bar{A} D^{-1} B \big).
\end{align}

\begin{figure}[ht!]
\centering
\includegraphics[scale=0.55]{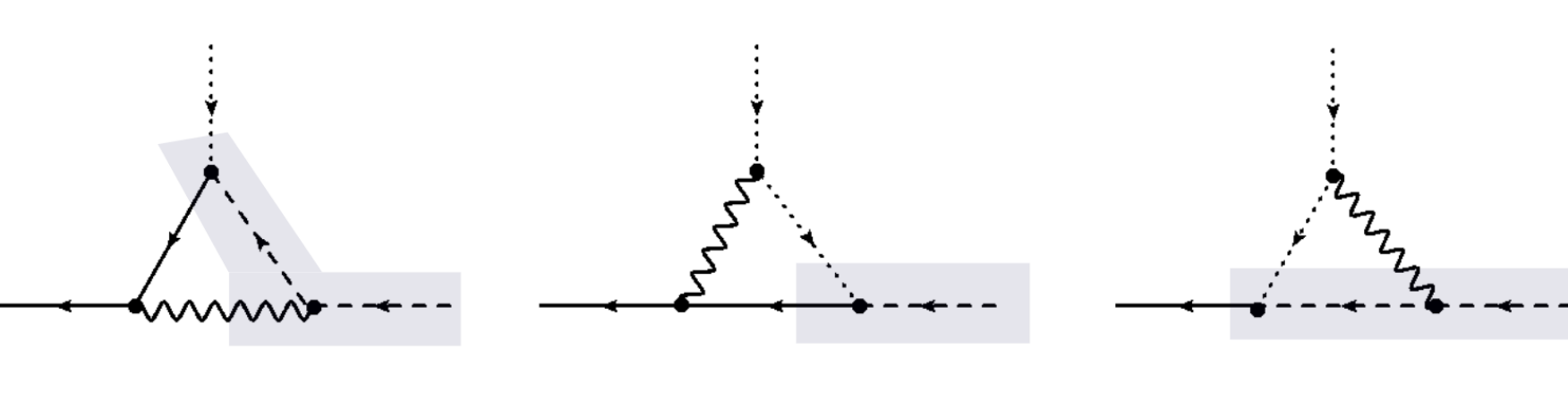}
\caption{Redrawn one-loop Feynman diagrams with a shaded area indicating the contribution of the ``effective vertex" appearing in Eq.~(\ref{lJ}).
}
\label{couplingYukawashaded}
\end{figure}

Green's functions with one external gluino field can be generated via functional differentiation with respect to the gluino source $J$ (cf. Eqs.~(\ref{PartFun}), (\ref{Seff})):
\begin{align}
    \label{lJ}
    \la(x): \quad \quad e^{-S'} & \rightarrow -\frac{d}{dJ_x} \, e^{-S'}|_{J=0} = \frac{1}{2} \, D^{-1}_{x,y} \, C^{-1} \, (\bar{A} + B')_y^T \, e^{-S'}|_{J=0} 
%    \bar{\la}: \quad \quad e^{-S'} & \rightarrow \frac{1}{2} \, (\bar{A} + B') \, D^{-1} \, e^{-S'}|_{J=0}
%    \label{lbarJ}
\end{align}
The above expression gives rise to all 3 diagrams of Fig.~\ref{couplingYukawa}; the diagrams are redrawn in Fig.~\ref{couplingYukawashaded} with a shaded area indicating the contribution of the ``effective vertex" 1/2  $D^{-1} C^{-1} \, (\bar{A} + B')^T$ appearing in Eq.~(\ref{lJ}) (note that $D$ contains contributions with zero or more gluons). We note also the factor of 1/2 present in Eq.~(\ref{lJ}); it is similar to the factor accompanying closed gluino loops, even though it does not stem from the Pfaffian.

In order to compute Green's functions with two external gluinos, for example $\la(x) \la(y)$, we have to consider the following second derivative with respect to the external source $J$:
\begin{align}
    \la(x) \la(y): \quad \quad  e^{-S'} & \rightarrow \bigg(-\frac{d}{dJ_x}\bigg) \bigg(-\frac{d}{dJ_y}\bigg) \, e^{-S'}|_{J=0} 
%    &= \bigg(-\frac{d}{dJ_x}\bigg) \, \frac{1}{2} \, \sum_z \,  M^{-1}_{y,z} \, (\bar{A} + B')_z^T \, e^{-S_{\text{eff}}}
\end{align}
Gluon fields contained in the matrices $M^{-1}$ and $D^{-1}$ of Eqs. (A7), (A8), can be extracted via a series expansion in g; thus, one gluon field emerges by calculating the quantity $g \frac{\partial}{\partial g}(M^{-1})\Bigr|_{g=0}$:
\begin{align}
%    & g \frac{\partial}{\partial g} (MM^{-1})\Bigr|_{g=0} =0 \nonumber \\
    & g \frac{\partial}{\partial g} (M^{-1})\Bigr|_{g=0} = -M^{-1} \bigg( g \frac{\partial M}{\partial g} \bigg) M^{-1} \Bigr|_{g=0}
\end{align}
where $g \frac{\partial M}{\partial g}$ is the normal vertex with two gluino fields and one gluon field. Similarly, extraction of two gluon fields follows from:
\begin{align}
     \frac{1}{2} g^2 \frac{\partial^2}{\partial g^2} (M^{-1})\Bigr|_{g=0} & = - \frac{1}{2} \, g^2 \, \frac{\partial}{\partial g} \bigg( M^{-1} \frac{\partial M}{\partial g} M^{-1} \bigg)\Bigr|_{g=0}  \nonumber \\
    & = g^2 M^{-1} \bigg(\frac{\partial M}{\partial g} \bigg) M^{-1} \bigg( \frac{\partial M}{\partial g} \bigg) M^{-1} \Bigr|_{g=0} - \frac{1}{2} \, g^2 M^{-1} \, \frac{\partial^2 M}{\partial g^2} \, M^{-1}
\end{align}

The term with $\frac{\partial^2 M}{\partial g^2}$ appears only on the lattice.

\begin{acknowledgements}
 This work was co-funded by the European Regional Development Fund and the Republic of Cyprus through the Research and Innovation Foundation (Project: EXCELLENCE/0421/0025). M.C. also acknowledges partial support from the Cyprus University of Technology under the ``POST-DOCTORAL" programme.
 
 We would like to acknowledge the insightful discussions we had with Haralambos Panagopoulos, which enriched our work. We also thank Georg Bergner for fruitful discussions and helpful comments.
\end{acknowledgements}


\begin{thebibliography}{99}
\bibitem{Bertone:2004pz}
G.~Bertone, D.~Hooper and J.~Silk,
``Particle dark matter: Evidence, candidates and constraints'',
Phys. Rept. \textbf{405} (2005), 279-390
doi:10.1016/j.physrep.2004.08.031
[arXiv:hep-ph/0404175 [hep-ph]].

\bibitem{Ellis:1983ew}
J.~R.~Ellis, J.~S.~Hagelin, D.~V.~Nanopoulos, K.~A.~Olive and M.~Srednicki,
``Supersymmetric Relics from the Big Bang'',
Nucl. Phys. B \textbf{238} (1984), 453-476
doi:10.1016/0550-3213(84)90461-9

\bibitem{Susskind:1982mw}
L.~Susskind,
``The Gauge Hierarchy problem, Technicolor, Supersymmetry, and all that'',
Phys. Rept. \textbf{104} (1984), 181-193
doi:10.1016/0370-1573(84)90208-4

\bibitem{ParticleDataGroup:2022pth}
R.~L.~Workman \textit{et al.} [Particle Data Group],
``Review of Particle Physics'',
PTEP \textbf{2022} (2022), 083C01
doi:10.1093/ptep/ptac097

\bibitem{Schaich:2022xgy}
D.~Schaich,
``Lattice studies of supersymmetric gauge theories'',
Eur. Phys. J. ST \textbf{232} (2023) no.3, 305-320
doi:10.1140/epjs/s11734-022-00708-1
[arXiv:2208.03580 [hep-lat]].

\bibitem{Giedt:2009yd}
J.~Giedt,
``Progress in four-dimensional lattice supersymmetry'',
Int. J. Mod. Phys. A \textbf{24} (2009), 4045-4095
doi:10.1142/S0217751X09045492
[arXiv:0903.2443 [hep-lat]].

\bibitem{Schaich:2014}
D. Schaich,
``Progress and prospects of lattice supersymmetry'',
PoS (\textbf{LATTICE2018}) 005
[arXiv:1810.09282].

\bibitem{Bergner:2016sbv}
G.~Bergner and S.~Catterall,
``Supersymmetry on the lattice'',
Int. J. Mod. Phys. A \textbf{31} (2016) no.22, 1643005
doi:10.1142/S0217751X16430053
[arXiv:1603.04478 [hep-lat]].

\bibitem{Athron:2007ry}
P.~Athron and D.~J.~Miller,
``A New Measure of Fine Tuning'',
Phys. Rev. D \textbf{76} (2007), 075010
doi:10.1103/PhysRevD.76.075010
[arXiv:0705.2241 [hep-ph]].

\bibitem{Montvay:2002}
F. Farchioni, C. Gebert, R. Kirchner, I. Montvay, A. Feo, G. Munster, T. Galla and A. Vladikas,
``The supersymmetric Ward identities on the lattice'',
Eur. Phys. J. D \textbf{76} (2002), 719
[arXiv:hep-lat/0111008].

\bibitem{Ali:2018fbq}S.~Ali, H.~Gerber, I.~Montvay, G.~M\"unster, S.~Piemonte, P.~Scior, G.~Bergner,
``Analysis of Ward identities in supersymmetric Yang-Mills theory'',
Eur.\ Phys.\ J.\ C {\bf 78} (2018) 404
  doi:10.1140/epjc/s10052-018-5887-9
[arXiv:1802.07067 [hep-lat]].

\bibitem{Taniguchi:1999fc}
Y.~Taniguchi,
``One loop calculation of SUSY Ward-Takahashi identity on lattice with Wilson fermion'',
Phys. Rev. D \textbf{63} (2000), 014502
doi:10.1103/PhysRevD.63.014502
[arXiv:hep-lat/9906026 [hep-lat]].

\bibitem{Curci:1986sm}
G.~Curci and G.~Veneziano,
``Supersymmetry and the Lattice: A Reconciliation?'',
Nucl. Phys. B \textbf{292} (1987), 555-572
doi:10.1016/0550-3213(87)90660-2

\bibitem{Sugino:2008yp}
F.~Sugino,
``Lattice Formulation of Two-Dimensional N=(2,2) SQCD with Exact Supersymmetry'',
Nucl. Phys. B \textbf{808} (2009), 292-325
doi:10.1016/j.nuclphysb.2008.09.035
[arXiv:0807.2683 [hep-lat]].

\bibitem{Catterall:2009it}
S.~Catterall, D.~B.~Kaplan and M.~Unsal,
``Exact lattice supersymmetry'',
Phys. Rept. \textbf{484} (2009), 71-130
doi:10.1016/j.physrep.2009.09.001
[arXiv:0903.4881 [hep-lat]].

\bibitem{Giedt:2014}
S. Catterall, D. Schaich, P.H. Damgaard, T. DeGrand and J.~Giedt,
``N = 4 supersymmetry on a space-time lattice'',
Phys. Rev. D \textbf{90} (2014), 065013
[arXiv:1405.0644].

\bibitem{Jones:1983ip}
D.~R.~T.~Jones,
``More on the Axial Anomaly in Supersymmetric {Yang-Mills} Theory'',
Phys. Lett. B \textbf{123} (1983), 45-46
doi:10.1016/0370-2693(83)90955-3

\bibitem{Wess:1992cp}
J.~Wess and J.~Bagger,
``Supersymmetry and supergravity'',
Princeton University Press, 1992,
ISBN 978-0-691-02530-8

\bibitem{Martin:1997ns}
S.~P.~Martin,
``A Supersymmetry primer'',
Adv. Ser. Direct. High Energy Phys. \textbf{18} (1998), 1-98
doi:10.1142/9789812839657\_0001
[arXiv:hep-ph/9709356 [hep-ph]].

\bibitem{Costa:2017rht}
M.~Costa and H.~Panagopoulos,
``Supersymmetric QCD on the Lattice: An Exploratory Study'',
Phys. Rev. D \textbf{96} (2017) no.3, 034507
doi:10.1103/PhysRevD.96.034507
[arXiv:1706.05222 [hep-lat]].

\bibitem{Costa:2018mvb}
M.~Costa and H.~Panagopoulos,
``Supersymmetric QCD: Renormalization and Mixing of Composite Operators'',
Phys. Rev. D \textbf{99} (2019) no.7, 074512
doi:10.1103/PhysRevD.99.074512
[arXiv:1812.06770 [hep-lat]].

\bibitem{Herodotou:2022xhz}
H.~Herodotou, M.~Costa and H.~Panagopoulos,
``Fine-Tuning of the Yukawa and Quartic Couplings in Supersymmetric QCD'',
PoS \textbf{LATTICE2022} (2022), 276 
doi:10.22323/1.430.0276
[arXiv:2210.03695 [hep-lat]].

\bibitem{Herodotou:2023xhz}
M.~Costa, H.~Herodotou and H.~Panagopoulos,
``Renormalization of the Yukawa and Quartic Couplings in $\mathcal{N} = 1$ Supersymmetric QCD'',
PoS \textbf{LATTICE2023} (2023), to appear.

\bibitem{Costa:2020keq}
M.~Costa, H.~Herodotou, P.~Philippides and H.~Panagopoulos,
``Renormalization and mixing of the Gluino-Glue operator on the lattice'',
Eur. Phys. J. C \textbf{81}, no.5, 401 (2021)
doi:10.1140/epjc/s10052-021-09173-x
[arXiv:2010.02683 [hep-lat]].

\bibitem{Costa:2021pfu}
M.~Costa, G.~Panagopoulos, H.~Panagopoulos and G.~Spanoudes,
``Gauge-invariant Renormalization of the Gluino-Glue operator'',
Phys. Lett. B \textbf{816}, 136225 (2021)
doi:10.1016/j.physletb.2021.136225
[arXiv:2102.02036 [hep-lat]].

\bibitem{Bergner:2022wnb}
G.~Bergner, M.~Costa, H.~Panagopoulos, I.~Soler and G.~Spanoudes,
``Perturbative renormalization of the supercurrent operator in lattice N=1 supersymmetric Yang-Mills theory'',
Phys. Rev. D \textbf{106}, no.3, 034502 (2022)
doi:10.1103/PhysRevD.106.034502
[arXiv:2205.02012 [hep-lat]].

\bibitem{Larin:1993tq}
S.~A.~Larin,
``The Renormalization of the axial anomaly in dimensional regularization'',
Phys. Lett. B \textbf{303} (1993), 113-118
doi:10.1016/0370-2693(93)90053-K
[arXiv:hep-ph/9302240 [hep-ph] containing an extra section].

\bibitem{Chanowitz:1979zu}
M.~S.~Chanowitz, M.~Furman and I.~Hinchliffe,
``The Axial Current in Dimensional Regularization'',
Nucl. Phys. B \textbf{159} (1979), 225-243
doi:10.1016/0550-3213(79)90333-X

\bibitem{tHooft:1972}
G.~'t Hooft and M.~J.~G.~Veltman,
``Regularization and Renormalization of Gauge Fields'',
Nucl. Phys. B \textbf{44} (1972), 189-213
doi:10.1016/0550-3213(72)90279-9

\bibitem{Siegel:1979}
W.~Siegel,
``Supersymmetric Dimensional Regularization via Dimensional Reduction'',
Phys. Lett. B \textbf{84} (1979), 193-196
doi:10.1016/0370-2693(79)90282-X

\bibitem{Patel}
A.~Patel and S.~R.~Sharpe,
``Perturbative corrections for staggered fermion bilinears'',
Nucl. Phys. B \textbf{395} (1993), 701-732
doi:10.1016/0550-3213(93)90054-S
[arXiv:hep-lat/9210039 [hep-lat]].

\bibitem{Buras:1989xd} A.~J.~Buras, P.~H.~Weisz, 
``QCD nonleading corrections to weak decays in dimensional regularization and 't Hooft-Veltman schemes'', Nucl. Phys. {\bf B333} (1990) 66.

\bibitem{quartic:2023}
M.~Costa, H.~Herodotou and H.~Panagopoulos,
``Supersymmetric QCD on the Lattice: Fine-Tuning and Counterterms for the Quartic Couplings'', in preparation.

\bibitem{Chetyrkin:1999pq}
K.~G.~Chetyrkin and A.~Retey,
``Renormalization and running of quark mass and field in the regularization invariant and MS-bar schemes at three loops and four loops'',
Nucl. Phys. B \textbf{583} (2000), 3-34
doi:10.1016/S0550-3213(00)00331-X
[arXiv:hep-ph/9910332 [hep-ph]].

\bibitem{Ruijl:2017eht}
B.~Ruijl, T.~Ueda, J.~A.~M.~Vermaseren and A.~Vogt,
``Four-loop QCD propagators and vertices with one vanishing external momentum'',
JHEP \textbf{06} (2017), 040
doi:10.1007/JHEP06(2017)040
[arXiv:1703.08532 [hep-ph]].

\bibitem{Luthe:2017ttg}
T.~Luthe, A.~Maier, P.~Marquard and Y.~Schroder,
``The five-loop Beta function for a general gauge group and anomalous dimensions beyond Feynman gauge'',
JHEP \textbf{10} (2017), 166
doi:10.1007/JHEP10(2017)166
[arXiv:1709.07718 [hep-ph]].

\bibitem{Chetyrkin:2017bjc}
K.~G.~Chetyrkin, G.~Falcioni, F.~Herzog and J.~A.~M.~Vermaseren,
``Five-loop renormalisation of QCD in covariant gauges'',
JHEP \textbf{10} (2017), 179
doi:10.1007/JHEP10(2017)179
[arXiv:1709.08541 [hep-ph]].

\bibitem{Gracey:2018fkg}
J.~A.~Gracey and R.~M.~Simms,
``Renormalization of QCD in the interpolating momentum subtraction scheme at three loops'',
Phys. Rev. D \textbf{97} (2018) no.8, 085016
doi:10.1103/PhysRevD.97.085016
[arXiv:1801.10415 [hep-th]].

\bibitem{Baikov:2019zmy}
P.~A.~Baikov and K.~G.~Chetyrkin,
``Transcendental structure of multiloop massless correlators and anomalous dimensions'',
JHEP \textbf{10} (2019), 190
doi:10.1007/JHEP10(2019)190
[arXiv:1908.03012 [hep-ph]].

\bibitem{Wellegehausen:2018opt}
B.~Wellegehausen and A.~Wipf,
``$\mathcal{N}=1$ Supersymmetric $SU(3)$ Gauge Theory - Towards simulations of Super-QCD'',
PoS \textbf{LATTICE2018} (2018), 210
doi:10.22323/1.334.0210
[arXiv:1811.01784 [hep-lat]].

\bibitem{Donini:1997hh}
A.~Donini, M.~Guagnelli, P.~Hernandez and A.~Vladikas,
``Towards N=1 superYang-Mills on the lattice'',
Nucl. Phys. B \textbf{523} (1998), 529-552
doi:10.1016/S0550-3213(98)00166-7
[arXiv:hep-lat/9710065 [hep-lat]].
\end{thebibliography}
\end{document}